# Review of Strong Field Approximation and Investigating Semiclassical Evolution Approach

Renjing Xu[1], Donghao Zhang[1], Anatoli Kheifet[2] and Igor Ivanov[2]

[1]Australian National University, College of Engineering and Computer Science
Canberra, ACT, 0200, Australia

[2]Research School of Physical Sciences, The Australian National University, Canberra ACT 0200, Australia
e-mail address: Igor.Ivanov@anu.edu.au

This paper theoretically analyzes the behavior of an atom driven by a strong electro-magnetic field. Moreover, besides traditional quantum mechanics method, we also investigate semiclassical approaches to this problem. We first performed strong field approximation for system of an atom driven by a strong electromagnetic field in velocity gauge. Our simulation result is consistent with theories and close to experiments except some reasonable difference caused by different parameters and omitted bound and final states in the transition amplitude. Next, a new semiclassical approach is used to solve Volkov wave function. We prove that semiclassical approximation works well in predicting the particle evolution in quantum world, especially for system in a strong electromagnetic field with low frequency. Finally, we also briefly illustrated how to use semiclassical approximation to get the same results as strong field approximation and also partly tested the viability of the semiclassical approach to Teller Potential. This part needs future work to accomplish. But still, semiclassical approximation provides a potentially new method to solve complicated system, which might be more effective than traditional quantum mechanics recipe.

**Keywords:** SFA, Semiclassical

## I. INTRODUCTION

In recent years, the laser physics grows fast; intense laser fields have become widely-used, with a wide frequency range, in the form of short even ultrafast pulses. When the laser field are strong enough to compete with Coulomb forces for determining the state of the atomic system; then, the role of higher order processes besides ordinary single-photon absorption or emission becomes significant. Strong Field Approximation is a widely-known method used to simplify and analyse those high order processes, such as: multiphoton ionization, harmonic generation and laser-assisted electron-atom collisions [1]. SFA assumes: the laser field would not change the initial bound state of the atoms; after the electrons are ionized to the continuum, they will not feel the presence of their parent atoms [2, 3].

Semiclassical approximation is another approach to describe the behaviour of atom driven by strong electro-magnetic field. This approximation to quantum mechanics is available in the situation that the de Broglie wavelength $\lambda = \frac{\hbar}{p}$ of a particle with momentum $p$ is greatly shorter than the length scales across which the potential of the system changes largely. In such short wavelength assumption, the particle could be regarded as a point and bounce off potential walls, which is the same as it does in classical mechanics. In quantum world, the particle could interference with different version of itself evolving along different classical trajectories, which is impossible in classical mechanics. The semiclassical approximation is made by taking the limit $\hbar \to 0$ in quantum mechanics to restrict the quantum quantities to follow their classical counterpart [4].

## II. THEOREY

Basic Theories & Equations
Gauge [5]:

$$\mathcal{E} = -\nabla \phi - \frac{\partial A}{\partial t} \quad (1)$$

$$\mathcal{B} = \nabla \times A \quad (2)$$

Where $\mathcal{E}$ is electric field, $\mathcal{B}$ is the magnetic field, A is vector potential and ϕ is scalar potential (by using Maxwell equation, A and ϕ also satisfy homogeneous wave equation as $\mathcal{E}$ and $\mathcal{B}$).

A and ϕ are not uniquely defined by Eq.1 and Eq2, $\mathcal{E}$ and $\mathcal{B}$ will be invariant when following relations are met, where Λ is an arbitrary real, differentiable function of r and t :

$$A' = A + \nabla \quad (3)$$

$$\phi' = \phi - \frac{\partial \Lambda}{\partial t} \quad (4)$$

Now, A is transformed to A' and ϕ is transformed to ϕ' by above relation. Such process is called Gauge Transformation. The freedom of choosing Λ leads to Gauge Invariance.

Coulomb gauge is a widely-used gauge for potentials, which is defined by following condition (given by choosing appropriate Λ):

$$\nabla \cdot A = 0 \qquad (5)$$

Coulomb gauge is frequently used when there is no source, hence, $\phi = 0$. Then the electric field is given as:

$$\mathcal{E} = -\frac{\partial A}{\partial t} \qquad (6)$$

Time-Dependent Schrodinger equation (TDSE) for Dipole Approximation [6]:

Dipole Approximation is valid when following conditions met:

Wavelength of the laser field is large compared with the size of the atomic system.

The laser intensity is not high enough to induce relativistic effect.

These two conditions imply that the spatial variation of electric field is negligible i.e. $\mathbf{A}(\mathbf{r}_0, t) = \mathbf{A}(t)$. Under such condition, $\mathcal{B} = \nabla \times \mathbf{A}$ vanishes and hence $\mathcal{E} = -\frac{\partial \mathbf{A}(t)}{\partial t}$. So, choosing Coulomb Gauge condition, the Hamiltonian of the N-electron atom system comprises two components:

$$H(t) = H_0 + H_{int}(t) \qquad (7)$$

$$H_0 = \frac{\hbar}{2m}\sum_{i=1}^{N}\nabla_{r_i}^2 - \sum_{i=1}^{N}\frac{Ze^2}{(4\pi\epsilon_0)r_i} + \sum_{i<j=1}^{N}\frac{e^2}{(4\pi\epsilon_0)r_{ij}} \qquad (8)$$

$$H_{int}(t) = \frac{e}{m}\mathbf{A}(t) \cdot \widehat{\mathbf{P}} + \frac{e^2 N}{2m}\mathbf{A}^2(t) \qquad (9)$$

Where $H_{int}(t)$ is interaction Hamiltonian, $\mathbf{P} = \sum_{i=1}^{N}\widehat{\mathbf{p}}_i$ ($\widehat{\mathbf{p}}_i$ is the momentum operator), e is the electric charge of an electron = 1.602*10^(-19) coulombs, Z is the atomic number, m is the mass of an electron = 9.10938291*10^(-31) kilograms, $r_{ij} = |\mathbf{r}_i - \mathbf{r}_j|$ and $\epsilon_0$ is the electric permittivity of free space.

Note the p here (the eignvalue of $\widehat{p}_i$) is not the common momentum we usually use (p = mv), it is called canonical momentum, which has following relation with v for the electrons in laser field:

$$p(t) = mv(t) - eA(r_0, t) \qquad (10)$$

Hence, the corresponding TDSE for an N-electron atom system under electromagnetic field is:

$$i\hbar\frac{\partial}{\partial t}\Psi(X,t) = [H_0 + \frac{e}{m}\mathbf{A}(t) \cdot \widehat{\mathbf{P}} + \frac{e^2 N}{2m}\mathbf{A}^2(t)]\Psi(X,t) \qquad (11)$$

Where $\Psi(X, t)$ is the wave function which describing the state of the atom system and $X \equiv (q_1, q_2, q_3, \ldots, q_N)$ which denotes the series of the coordinates of the N electrons and $q_i \equiv (r_i, \sigma_i)$ are the space and spin coordinates of the corresponding electrons.

Choice of Gauge [7]:

Under the gauge transformation in the 1st part, the transfer of the wave function is shown below:

$$\Psi' = e^{-\frac{ie}{\hbar}\Lambda}\Psi \qquad (12)$$

where $\Psi'$ satisfies following equation (not necessary in Coulomb Gauge):

$$i\hbar\frac{\partial}{\partial t}\Psi(r,t) = [H_0 - i\hbar\frac{e}{m}(\mathbf{A}' \cdot \nabla + \nabla \cdot \mathbf{A}'(t)\frac{e^2 N}{2m}\mathbf{A}'^2(t) - e\phi']\Psi(r,t) \qquad (13)$$

Now, the gauge transformation is performed as following:

$$\Lambda = -\mathbf{A}(t) \cdot \mathbf{R} \qquad (14)$$

where $= \sum_{i=1}^{N} r_i$, which is the sum of the coordinates of the N electrons.

Then, after Gauge Transformation, we have:

$$\mathbf{A}' = \mathbf{0} \qquad (15)$$

$$\phi = -\mathcal{E}(t) \cdot \mathbf{R} \qquad (16)$$

$$\Psi^L(\mathbf{X}, t) = e^{\frac{ie}{\hbar}\mathbf{A}(t) \cdot \mathbf{R}}\Psi(\mathbf{X}, t) \qquad (17)$$

Hence, the corresponding TDSE for new wave function

$$i\hbar\frac{\partial}{\partial t}\Psi^L(\mathbf{X}, t) = [H_0 + e\mathcal{E}(t) \cdot \mathbf{R}]\Psi^L(\mathbf{X}, t) \qquad (18)$$

This is referred as Length Gauge, as the corresponding interaction Hamiltonian is:

$$H_{int}^L = e\mathcal{E}(t) \cdot \mathbf{R} \qquad (19)$$

Recalled the coulomb gauge:

$$H^C_{int}(t) = \frac{e}{m}\mathbf{A}(t) \cdot \widehat{\mathbf{P}} + \frac{e^2 N}{2m}\mathbf{A}^2(t) \qquad (20)$$

The Velocity Gauge is the same as the coulomb gauge:

$$i\hbar\frac{\partial}{\partial t}\Psi^V(\mathbf{X}, t) = \left[H_0 + \frac{e}{m}\mathbf{A}(t) \cdot \widehat{\mathbf{P}}\frac{e^2}{2m}\mathbf{A}^2(t)\right]\Psi^V(\mathbf{X}, t) \qquad (21)$$

which is referred as Velocity Gauge, since the corresponding interaction Hamiltonian is:

$$H_{int}^V = \frac{e}{m}\mathbf{A}(t)\cdot\hat{\mathbf{P}} + \frac{e^2}{2m}\mathbf{A}^2(t) \quad (22)$$

The two different Gauges are widely-used in different fields of Strong Field Approximations. Using which gauge to do strong field approximation is being a debate for a long time. In Ref. [8], it illustrates that the calculation by using velocity gauge is expected to be more effective in higher-order harmonic generation. In Ref. [9], the length gauge is suggested to be used in describing the interference effect for intense-field-molecule interaction. In Ref. [10], length gauge has more advantages in simulating above-threshold ionization of atoms and molecules and nonsequential double ionization. Actually, the difference of the simulation results from using different gauge arises from the assumptions of the approximation in different conditions. When the TDSE is solved appropriately, there will be no difference between these two gauges. In this report, the TDSE will solved by using Velocity Gauge.

Gordon-Volkov wave function [11]

The Gordon-Volkov (or Volkov) wave function is an exact solution of the TDSE for describing the motion of a free, charged particle in a plane-wave electromagnetic field. Volkov wave function has a plenty of applications in calculate ionization of atoms [12], excitation in band-gap semiconductors [13] and scattering of charged particle [14]. All the above applications could be improved by introduction of appropriate corrections to the Volkov solution because of the presence of the atomic potential besides the plane-wave field [15]. Hence, understating Volkov wave is essential for solving more complex problem, such as my topic 'behaviour of an atom driven by strong electric field'.

Under the velocity gauge and dipole approximation, TDSE is shown below:

$$i\hbar\frac{\partial}{\partial x}|\psi(t)\rangle = [-\frac{\hbar}{2m}\nabla^2 - \frac{ei\hbar}{m}\mathbf{A}(t)\cdot\nabla + \frac{e^2}{2m}\mathbf{A}^2(t)]|\psi(t)\rangle \quad (23)$$

Change the state vector by taking $\tilde{\psi}(t,p) = \langle p|\psi(t)\rangle$:

$$i\hbar\frac{\partial}{\partial x}\tilde{\psi}(t,\mathbf{p}) = [\frac{\mathbf{p}^2}{2m} + \frac{e}{m}\mathbf{A}(t)\cdot\mathbf{p} + \frac{e^2}{2m}\mathbf{A}^2(t)]\tilde{\psi}(t,\mathbf{p}) \quad (24)$$

By using Fourier transform to solve this equation:

$$\tilde{\psi}(t,p) = e^{-\frac{i}{2m}\int_0^t[p+eA(\tau)]^2 d\tau} \quad (25)$$

The inverse Fourier transform of Eq.25 is taken, then, the solution in r space is given as following:

$$\psi_p(t,r) = Ce^{p\cdot r - \frac{i}{2m}\int_0^t[p+eA(\tau)] d\tau} \quad (26)$$

Where C is an arbitrary constant right now

To find C, the normalization condition needs to meet:

$$\langle\psi_p(t,r)|\psi_{p'}(t,r)\rangle = \delta(p-p') \quad (27)$$

C is found to be $(2\pi)^{\frac{3}{2}}$

Finally, the Volkov wave function under velocity gauge and dipole approximation is:

$$\psi_p(t,r) = (2\pi)^{\frac{3}{2}}e^{ip\cdot r - \frac{i}{2m}\int_0^t[p+eA(\tau)]^2 d\tau} \quad (28)$$

The same equation will be deduced by another method (semiclassical evolution) later.

Semiclassical Evolution [4]

For a 1–degree of freedom particle moving in a mildly varying potential, the free particle function to a wave function can be generalized to a wave function:

Where q is the canonical coordinate, $A(q,t)$ is slowing varying amplitude and $R(q,t)$ is fast varying phase. $A(q,t)$ and $R(q,t)$ are both real functions.

Now, the A and R here will be found by using semiclassical evolution to relate the classical mechanics and quantum mechanics.

Obviously, the wave function needs to satisfy TDSE:

$$\left(i\hbar\frac{\partial}{\partial t} + \frac{\hbar^2}{2m}\frac{\partial^2}{\partial q^2} - V(q)\right)\psi(q,t) = 0 \quad (30)$$

Note that the TDSE here could be regard as length gauge.

Substitute $\psi(q,t)$ in Eq. 29 into Eq.30, and separate imaginary and real parts, the following two equations are given:

The real part:

$$\frac{\partial R}{\partial t} + \frac{1}{2m}\left(\frac{\partial R}{\partial q}\right)^2 + V(q) - \frac{\hbar^2}{2mA}\frac{\partial^2}{\partial q^2}A = 0 \quad (31)$$

The imaginary part:

$$\frac{\partial A}{\partial t} + \frac{1}{m}\sum_{i=1}^{D}\frac{\partial A}{\partial q_i}\frac{\partial R}{\partial q_i} + \frac{1}{2m}A\frac{\partial^2 R}{\partial q^2} = 0 \quad (32)$$

As mentioned previously, in semiclassical evolution, $\hbar \to 0$. Hence, the term $\frac{\hbar^2}{2mA}\frac{\partial^2}{\partial q^2}A \to 0$ is dropped. Then, Eq. 31 becomes:

$$\frac{\partial R}{\partial t} + H\left(q,\frac{\partial R}{\partial t}\right) = 0 \quad (33)$$

This equation is known as the Hamilton-Jacobi equation in classical mechanics.

Now, in order to solve wave function $\psi(q,t)$ semiclassically, A and R need to be found.

For simplicity, following definition is given:

$$p_i = p_i(q,t) := \frac{\partial R}{\partial q_i} \quad (34)$$

By chain rule, we have:

$$dR = \frac{\partial R}{\partial t}dt + \frac{\partial R}{\partial q}dq \quad (35)$$

The relation could be illustrated by figure below:

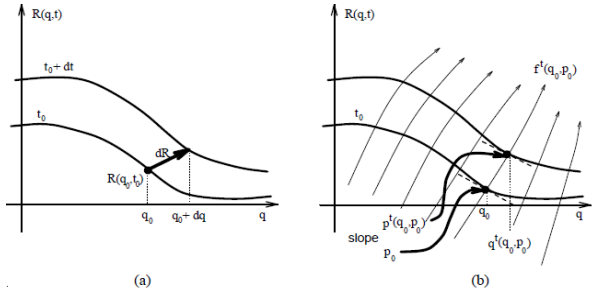

Figure1. Phase R(q,t) as function of q and t

Divide Eq35 by dt and then substitute Eq.33, the derivative of R(q,t) with respect to as yet arbitrary $\dot{q}$ is:

$$\frac{dR}{dt}(q,\dot{q},t) = -H(q,p) + \dot{q}\cdot p \quad (36)$$

Eq.36 can be integral and find R, however, before we can do this, yet unknown $\dot{q}$ needs to be determined.

Note that the p here is also a function of q and t. Use chain rule again we will have:

$$dp = d\frac{\partial R}{\partial q} = \frac{\partial^2 R}{\partial q \partial t}dt + \frac{\partial^2 R}{\partial q^2}dq \quad (37)$$

After substituting to Eq.33:

$$dp = -\left(\frac{\partial H}{\partial q} + \frac{\partial H}{\partial p}\frac{\partial R}{\partial q}\right)dt + \frac{\partial p}{\partial q}dq \quad (38)$$

Divide by dt again:

$$\dot{p}(q,\dot{q},t) + \frac{\partial H}{\partial q} = \left(\dot{q} - \frac{\partial H}{\partial p}\right)\frac{\partial p}{\partial q} \quad (39)$$

Here, $\dot{p}$ depends both on $\dot{q}$ and $\frac{\partial p}{\partial q}$, which are all yet unknown. However, if the arbitrary $\dot{q}$ is chosen to be $\frac{\partial H}{\partial p}$ such that right hand side of Eq. 39 disappears. Now, the function R(q,t) can be solved by integrating following first order differential equation to determine the evolving trajectory (q(t),p(t)):

$$\dot{q} = \frac{\partial H(q,p)}{\partial p} \quad (40)$$

$$\dot{p} = -\frac{\partial H(q,p)}{\partial q} \quad (41)$$

The initial conditions are given as:

$$q(t_0) = q' \quad (42)$$

$$p(t_0) = p' = \frac{\partial R}{\partial q}(q',t_0) \quad (43)$$

Eq. 40 and Eq. 41 are well-known Hamilton's equation in classical mechanics.

Now, R(q,t) can be integrating Eq. 36 along the trajectory (q(t),p(t)) as:

$$R(q,t) = R(q',t_0) + R(q,t;q',t_0) \quad (44)$$

$$R(q,t;q',t_0) = \int_{t_0}^{t}[\dot{q}(\tau)\cdot p(\tau) - H(q(\tau),p(\tau))]d\tau \quad (45)$$

For simplicity, $t_0 = 0$, then we have:

$$R(q,q',t) = R(q,t;q',0) \quad (46)$$

$R(q,q',t)$ is well-known Hamiltonian's principle function.

Next step is to find how A varies.

Physical quantity density is introduced here:

$$\rho(q,t) := A^2 = \psi^*(q,t)\psi(q,t) \quad (47)$$

For small volume denoted by $d^D q$, the amount of hypothetical particles inside it is $\rho(q,t)d^D q$. Assume that there is only one solution for $\psi$, and the amount of hypothetical particle is conserved. Then, the following relation of the initial condition and the condition at the time t is satisfied:

$$\rho(q,t)d^D q = \rho(q',0)d^D q' \quad (48)$$

The relation is also briefly illustrated in following figure:

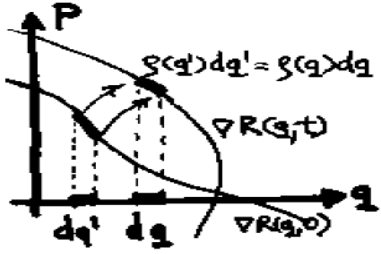

Figure2. Density Variation

The variation of the density could be found from Eq. 48 as:

$$\rho(q,t)d^D q = |\det \frac{\partial q'}{\partial q}|\rho(q',0) \quad (49)$$

Now, we have enough tools to figure out the evolved wave function in the semiclassical approximation:

$$\psi(q,t) = A(q,t)e^{\frac{iR(q,t)}{\hbar}} = \sqrt{\det\frac{\partial q'}{\partial q}} A(q,0)e^{\frac{i(R(q,0)+R(q,q',t))}{\hbar}}$$

$$= \sqrt{\det\frac{\partial q'}{\partial q}} e^{\frac{iR(q,q',t)}{\hbar}}\psi(q',0) \quad (50)$$

Later, this evolved wave function will be used to describe the behaviour of an atom or electron driven by a strong electromagnetic field, which does the same thing as strong field approximation and Volkov wave function.

## III. APPLY OF THEORY AND NUMERICAL RESULTS

Method, Result and Interpretation
Strong Field Approximation in Velocity Gauge
Now, I will focus on SFA for the system of an atom with one electron.
At the beginning, the Strong Field Approximation Amplitude (also known as transition amplitude) is introduced [2, 10]:

$$T = -\frac{i}{\hbar}\int_0^{T_i} u(T_i,\tau)V_{int}(\tau)u_0(\tau,0)d\tau \quad (51)$$

Where $u(T_i,\tau)$ is the evolution operator when the electron is ionized i.e. when the electron is described by Volkov wave function (as mentioned earlier), $u_0(\tau,0)$ is the evolution operator when the electron is in bound state and $V_{int}$ is the interaction energy ( $V_{int}(\tau)|\psi(t)\rangle = H_{int}(\tau)|\psi(t)\rangle$).

Here, I assume, the laser is switched off when the t < 0.

The TDSE for bound state:

$$\frac{\partial}{\partial t}|\psi(t)\rangle = H_0|\psi(t)\rangle \quad (52)$$

The time evolution of the wave function would be:

$$|\psi(t)\rangle = e^{\frac{iE_0 t}{\hbar}}|\psi(t_0)\rangle \quad (53)$$

Where E0 is the energy of the corresponding bound state
Hence, the evolution operator for an electron in bound state would be:

$$u_0(\tau,0) = e^{\frac{iE_0\tau}{\hbar}} \quad (54)$$

From Eq.28, the evolution operator for free electron under a strong electromagnetic field:

$$u(T_i,\tau) = e^{\frac{-i}{2m}\int_\tau^{T_i}[p\mp A(\tau)]^2 d\tau} \quad (55)$$

For velocity gauge, from Eq. 22, Hint is given for single electron as:

$$V_{int}^V(\tau) = \frac{e}{m}A(\tau)\cdot p + \frac{e^2}{2m}A^2(\tau) \quad (56)$$

Where p is the canonical momentum of the electron
Hence, combining these components, the transition amplitude is:

$$T = -\frac{i}{\hbar}\int_0^{T_i} e^{\frac{-i}{2m}\int_\tau^{T_i}[p+eA(\tau)]^2 d\tau}\left(\frac{e}{m}A(\tau)\cdot p + \frac{e^2}{2m}A^2(\tau)\right)e^{\frac{iE_0\tau}{\hbar}}d\tau \quad (57)$$

For simplicity, the atomic system of unit (e = 1, $\hbar = 1$, m = 1), also, for generalization, transition amplitude is not only work for electron, hence, the minus signs before e are deleted:

$$T = -i\int_0^{T_i} e^{\frac{-i}{2}\int_\tau^{T_i}[p+A(\tau)]^2 d\tau}\left(A(\tau)\cdot p + \frac{A^2(\tau)}{2}\right)e^{iE_0 t}d\tau \quad (58)$$

Note that transition amplitude is a quantity for certain direction. In order to get reasonable physical meaning, we need to integrate the square of the module of the transition amplitude over all direction for a given p. By using symmetric argument:

$$T_{total}(p) = \int_0^{2\pi}\int_0^{\pi} T(p)^* T(p)\sin(\theta)d\theta d\varphi \quad (59)$$

Where θ is the polar angle in spherical system, and φ is the azimuth angle

By taking $E = \frac{1}{2}v^2 = \frac{p^2}{2}$, transition amplitude can be related to energy:

$$T'(E) = T(p) \quad (60)$$

Then, Eq. 58 can be transferred to:

$$T'_{total}(E) = \int_0^{2\pi} \int_0^{\pi} T'(E)^* T'(E) \sin(\theta) d\theta d\varphi \quad (61)$$

Next step is to simulate Eq. 60 using FORTRAN, the corresponding parameters are:

$$A(\tau) = -\frac{\varepsilon_0}{\omega} \sin(\omega t)$$
$$\omega = 0.057$$
$$\varepsilon_0 = 0.1$$
$$E_0 = -0.5$$
$$T_1 = 4T$$
$$T = \frac{2\pi}{\omega}$$

The simulated plot is shown below:

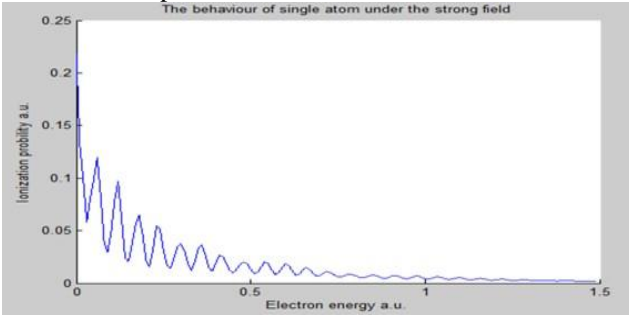

Figure3. The behaviour of single atom under the strong EM field

The experimental result of electron energy spectrum showing above-threshold ionization from reference is also presenting below [16]:

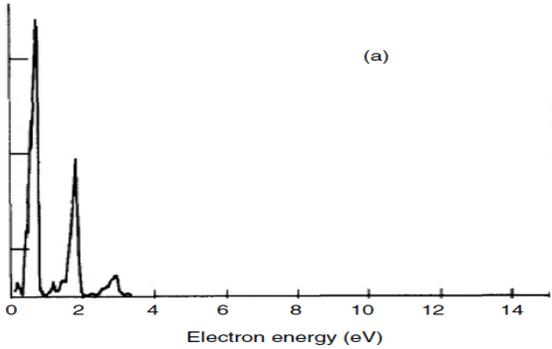

Figure4. Electron energy spectrum showing above-threshold ionization

My stimulated results is consistent with the experiment results as shown in Figure 4, which is showing that the n-photon ionization rate is proportional to $I^n$ (I the intensity of the electromagnetism wave, it is in atom unit and smaller than 1, thus, the ionization rate is decreasing when n is increasing.). However, they are not exactly the same. There are two reasons: 1. the parameters are different 2. the transition amplitude here is not exactly 'transition amplitude'.

The square of the module of the transition amplitude is corresponding to the ionization rate. The exact version of transition amplitude is:

$$f = \langle p|T|0\rangle \quad (62)$$
$$= \left\langle p \left| -\frac{i}{\hbar} \int_0^{T_i} u(T_i,\tau) V_{int}(\tau) u_0(\tau,0) d\tau \right| 0 \right\rangle$$

Where $|0\rangle$ is the bound state and $\langle p|$ is the bra version of the state of the free electron in the electromagnetic field.

Here, Eq. 62 is not the most important part while Eq. 51 is much more important, because we can free add bound state and final state into Eq.62 if T is given from Eq. 51. Also, comparing deriving Eq.58 and verifying Eq.58, verifying the exact transition amplitude is much easier, so I put more effort on finding and verifying Eq.58, and leave Eq.62 for future work.

Semiclassical Method for solving Volkov wave function
For simplicity, the derivation here is using atomic system of unit.
From Eq. 50, and for single electron, we have:

$$\psi(r,t) = e^{\frac{iR(q,q',t)}{\hbar}} \psi(r',0) \quad (63)$$

Note that $\sqrt{\det \frac{\partial q'}{\partial q}} = 1$, as I will illustrate it later.
As Eq. 45, we have:

$$R(r,r',t) = \int_0^t [\dot{r}(\tau) \cdot p(\tau) - H(r(\tau), p(\tau))] d\tau \quad (64)$$

$$p(\tau) = \dot{r}(\tau) - A(\tau) \quad (65)$$

$$\dot{r}(\tau) = p(\tau) + A(\tau) \quad (66)$$

$$H(r(\tau), p(\tau)) = \frac{1}{2}(p(\tau) + A(\tau))^2 \quad (67)$$

After substituting Eq.65 and Eq.66 into Eq.44, we have:

$$R = \int_0^t \left( p(\tau)(p(\tau) + A(\tau)) - \frac{(p(\tau) + A(\tau))^2}{2} \right) d\tau$$
$$= \int_0^t \left( \frac{p^2(\tau) - A^2(\tau)}{2} \right) d\tau \quad (68)$$

By integrating Eq.66, we also could relate the initial position and the position of the electron at time t:

$$r = \int_0^t (p(\tau) + A(\tau)) d\tau + r_i \quad (69)$$

Hence, we have:
$$r_i = r - \int_0^t (p(\tau) + A(\tau))d\tau \qquad (70)$$
Note that $\sqrt{\det \frac{\partial q'}{\partial q}} = \sqrt{\frac{\partial r_i}{\partial r}} = 1$

Use $\psi(r', 0) = e^{ipr'}$ as the testing wave function, and substitute it, Eq. 68 and Eq. 70 into Eq. 63:

$$\begin{aligned}\psi(r,t) &= e^{ip(\tau)(r-\int_0^t(p(\tau)+A(\tau))d\tau)}e^{i\int_0^t\left(\frac{p^2(\tau)-A^2(\tau)}{2}\right)d\tau} \\ &= e^{ip(\tau)r}e^{-i\int_0^t(\frac{p^2(\tau)-A^2(\tau)}{2}+p(\tau)A(\tau))d\tau} \\ &= e^{ip(\tau)\cdot r-\frac{i}{2}\int_0^t[p+A(\tau)]^2 d\tau}\end{aligned} \qquad (71)$$

Eq. 71 is exactly the same as Eq.28 which is the Volkov wave function calculated in quantum mechanics except there is no normalization factor and Eq. 71 is in atomic unit. Here, I prove that semiclassical approximation works well in predicting the particle evolution in quantum world, especially for system in strong electromagnetic field with low frequency.

*Semiclassical Method used in Strong Field Approximation* [4]

When it comes to the system of an atom with one electron driven by a strong electromagnetic field, the Hamiltonian needs to include the coulomb potential from the nuclei:
$$H = \frac{(p(\tau) + A(\tau))^2}{2} + V(\tau) \qquad (72)$$

Also, the $\sqrt{\det \frac{\partial q'}{\partial q}}$ may not be 1 as in the Volkov case, and the eignvalues of the $\det \frac{\partial q'}{\partial q}$ will change at each fold of the Lagrangian manifold crossing, which is shown below:

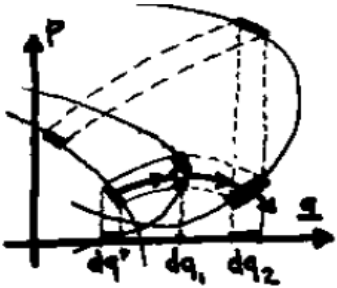

Figure5. Folding of the Lagrangian Surface

Hence, the sign of the Jacobian determinant can be tracked in following way:
$$\det \frac{\partial q'}{\partial q}\bigg|_j = e^{-i\pi m_j(q,q',t)}\left|\det \frac{\partial q'}{\partial q}\right|_j \qquad (73)$$

Hence, the new semiclassical approximation is corrected to:
$$\psi(q,t) = \int dq' \sum_j \sqrt{\det \frac{\partial q'}{\partial q}}\bigg|_j e^{\frac{iR_j(q,q',t)}{\hbar} - \frac{i\pi m_j(q,q',t)}{2}} \psi(q'_j, 0) \qquad (74)$$

Solve Eq. 74 for system of an atom with one electron under a strong electromagnetic field is much more complicated than system of free electron. This part needs future work to accomplish. But still, semiclassical approximation provides a potentially new method to solve complicated system, which might be more effective than traditional quantum mechanics recipe.

Apply semiclassical method to Teller potential
Teller potential is given as:
$$V(x) = -\text{sech}^2(x) \qquad (75)$$
The corresponding ground state potential is:
$$E_g = -0.5 \qquad (76)$$
In order to computer Eq.50, the q and p need to be solved by Eq.40 and Eq.41:
$$H = \frac{1}{2}(p + A) + V \qquad (77)$$
$$\dot{q} = \frac{\partial H(q,p)}{\partial p} = p + A \qquad (78)$$
$$\dot{p} = -\frac{\partial H(q,p)}{\partial q} = -2\tanh(x) * \text{sech}^2(x) \qquad (79)$$
Then, after converting Eq.50 to one dimension case, we have:
$$\psi(x,t) = \sqrt{\det \frac{\partial x}{\partial x'}} e^{ip_0(x-\int_0^t(p+A)dt)} e^{i\int_0^t\left(\frac{p^2-A^2}{2}-V\right)dt} \qquad (80)$$

Compare the semiclassical method for Teller potential to Volkov equation with increasing electric field

After coding, the results of the wave function for semiclassical method and Volkov equation are shown below (real part only):

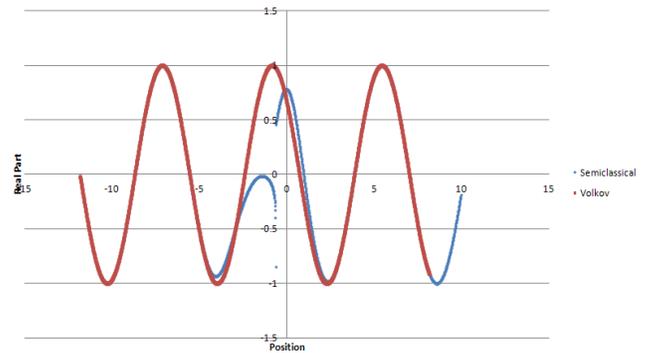

Figure6. For electric field Strength of 0.1

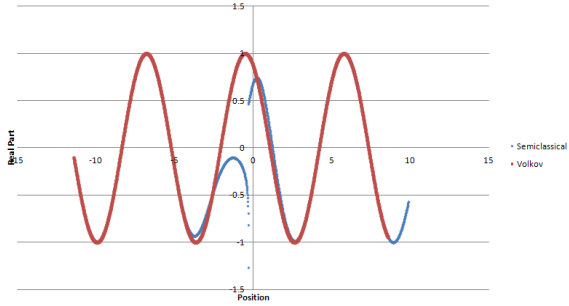

Figure7. For electric field Strength of 0.4

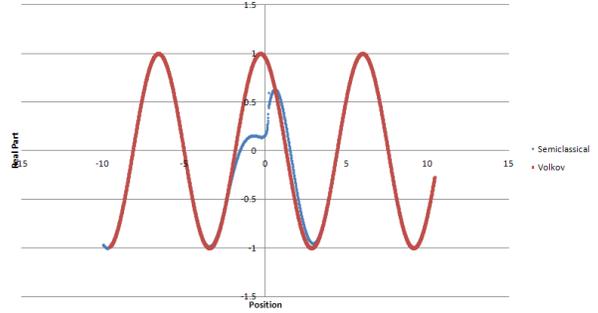

Figure11. For electric field Strength of 1.7

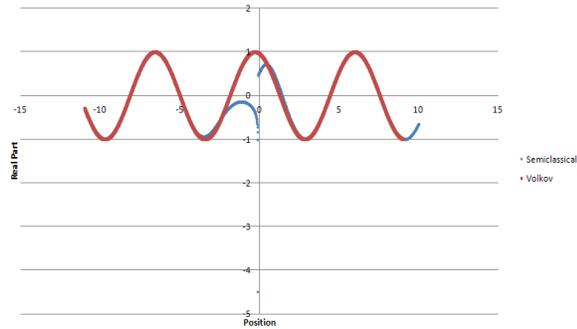

Figure8. For electric field Strength of 0.7

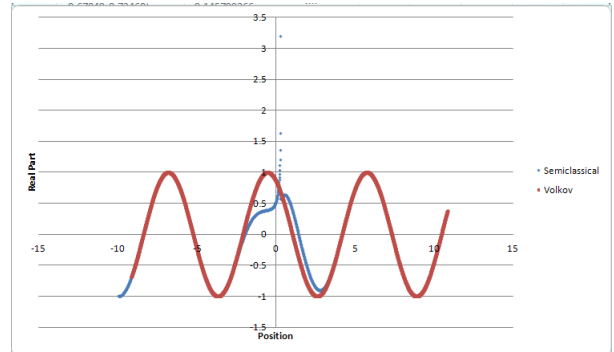

Figure12. For electric field Strength of 2.0

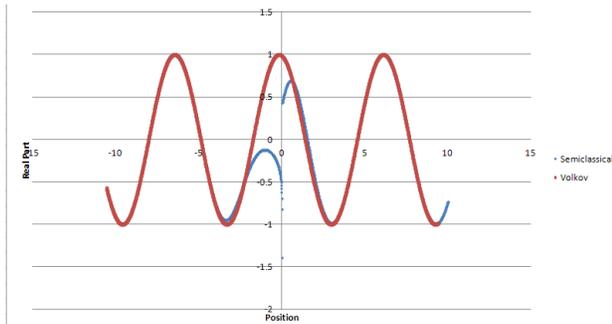

Figure9. For electric field Strength of 1

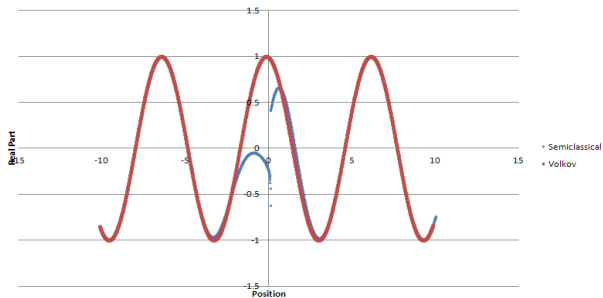

Figure10. For electric field Strength of 1.3

From figure 6 to 12, when the strength of electric field is increasing, the difference between the Volkov wave function and the semiclassical approach to the Teller Potential is decreasing as well. It is reasonable as that the influence from the Teller Potential is diminished with the increasing of the electric field strength. From above tests, the viability of the semiclassical approach is partly proved.

## IV. CONCLUSION

In this paper, we performed strong field approximation for system of an atom driven by a strong electromagnetic field in velocity gauge. Our simulation result is consistent with theories and close to experiments except some reasonable difference caused by different parameters and omitted bound and final states in the transition amplitude. Also, a new semiclassical approach is used to solve Volkov wave function. We prove that semiclassical approximation works well in predicting the particle evolution in quantum world, especially for system in strong electromagnetic field with low frequency. Finally, we also briefly illustrated how to use semiclassical approximation to get the same results as strong field approximation and also partly tested the viability of the semiclassical approach to Teller Potential. This part needs future work to accomplish. But still, semiclassical approximation provides a potentially new method to solve complicated system, which might be more effective than traditional quantum mechanics recipe.